\documentclass[preprint,12pt,authoryear]{elsarticle}
\usepackage{amsmath,amsfonts,amssymb,amsthm}
\usepackage[version=4]{mhchem}
\usepackage{chemfig}
\begin{document}

\begin{frontmatter}
\title{Benchmarking inscribed matter probes}

\author{Michael Hippke\corref{label3}}
\address{Sonneberg Observatory, Sternwartestr. 32, 96515 Sonneberg, Germany}
\ead{michael@hippke.org}

\author{Paul Leyland}
\address{Brnikat Ltd, 19a Hauxton Road, Little Shelford, Cambridge, CB22 5HJ, United Kingdom}
\ead{paul@brnikat.com}

\author{John G. Learned}
\address{High Energy Physics Group, Department of Physics and Astronomy, University of Hawaii, Manoa 327 Watanabe Hall, 2505 Correa Road Honolulu, Hawaii 96822 USA}
\ead{jgl@phys.hawaii.edu}

\begin{abstract}
We have explored the optimal frequency of interstellar photon communications and benchmarked other particles as information carriers in previous papers of this series. We now compare the latency and bandwidth of sending probes with inscribed matter. Durability requirements such as shields against dust and radiation, as well as data duplication, add negligible weight overhead at velocities $v<0.2\,$c. Probes may arrive in full, while most of a photon beam is lost to diffraction. Probes can be more energy efficient per bit, and can have higher bandwidth, compared to classical communication, unless a photon receiver is placed in a stellar gravitational lens. The probe's advantage dominates by order of magnitude for long distances (kpc) and low velocities ($<0.1\,$c) at the cost of higher latency.
\end{abstract}

\end{frontmatter}

\section{Introduction}
Sending a physical artifact can be the most energy-efficient choice for interstellar communications, because it can be done at almost arbitrarily low velocities, and thus low energies. An artifact can arrive at the destination in total, in contrast to a particle beam which is wider than the receiver in all realistic cases, so that most energy is lost. The obvious disadvantage is the large communication delay, e.g. sending a probe at a (relatively fast) $0.01\,c$ takes 438 years to the nearest star, Proxima.

While such communications have been described in the literature before \citep[e.g.,][]{2004Natur.431...47R}, they have never been rigorously examined in terms of data rate, bit per energy, latency, or durability. In this study, we benchmark the inscribed matter probes to optimal photon communication at keV energies \citep{2017arXiv171105761H}.

There are two main use cases for inscribed matter. The first is the placement of an artifact in a stellar system to be found at a later time, after the rise of a civilization and its exploration of the system. This scenario is similar to Arthur C. Clarke's moon monolith \citep{Clarke1953}. It was suggested to search the solar system for non-terrestrial artifacts \citep{1960Natur.186..670B,1995ASPC...74..425P,2004IAUS..213..487T,2012AcAau..72...15H}, particularly for starships \citep{Martin1980} in addition to classical SETI \citep{2016arXiv160904635G}. In our solar system, probes are speculated to be in geocentric, selenocentric, Earth-Moon libration, and Earth-Moon halo orbits \citep{1980Icar...42..442F,1983Icar...53..453V,1983Icar...55..337F}. Alternative ideas include the Kuiper belt \citep{2012AsBio..12..290L}, general technosignatures \citep{2017arXiv170407263W}, or even ``footprints of alien technology on Earth'' \citep{2012AcAau..73..250D}. A galactic library of knowledge would ease the need for fast communication, keeping the impatient Aliens busy while waiting for the first ``answer'' from the stars.

The second type is to establish a ``live'' communication between two distant species using one, or many, probes with onboard storage. In the case that the receiver needs to decelerate such a probe, it appears sensible to first establish contact using faster communication, e.g. using low data rate photons, perhaps even as a beacon. High data rate (high latency) probe exchange can follow after exchanging ephemerides and other protocol details \citep{2017arXiv171105761H}.

\section{Durability}
Space is not empty, but contains a low density of particles, mainly a plasma of hydrogen and helium atoms as well as electromagnetic radiation, magnetic fields, neutrinos, and dust. A probe traveling between the stars will be impacted and potentially damaged by these particles and photons. We now estimate the required shielding and remaining damage.

\subsection{Front shield}
Heads-on interaction with gas and dust in the interstellar medium at relativistic velocities erodes the front surface and produces craters due to explosive evaporation of surface atoms. Gas penetrates the front to a certain depth.

The erosion from dust impacts has been estimated as 0.1\,mm (0.5\,mm) for quartz material at $v=0.1\,c$ ($0.2\,c$) per pc \citep{2017ApJ...837....5H}. For the penetration depth of protons there are different estimates between 3\,mm in quartz or graphite \citep[at 0.2\,c,][]{2017ApJ...837....5H} and 10\,mm in titanium \citep[at 0.1\,c,][]{2009AcAau..64..644S}. Penetration depth is a strong function of velocity with a minimum requirement of 10\,m of titanium shielding at $v=0.995$.

Conservatively, the front shield should have a thickness of order $1 + 0.1d$\,cm where $d$ is the travel distance in pc (for $v=0.1\,$c). This assures that sufficient proton shielding remains after dust impact erosion towards the end of the journey.

\subsection{Side and bottom shields}
For dust impacting from the sides, lower velocities in the rest frame of the craft are expected (planetary, km\,s$^{-1}$) with typical grain sizes of order $\mu$m. This erosion has been determined on average as $22\,\mu$m$\,10^6\,{\rm yr}^{-1}$ for iron meteorites in the solar system \citep{1981P&SS...29.1109S} and 6 orders of magnitude lower in interstellar space \citep{doi:10.1146/annurev-astro-081309-130846}.

These low levels of erosion are dominated by single large impacts. The largest expected particles are of $\mu$m sizes, and it is unlikely ($<1\,$\%) for a spacecraft to be hit by dust $>5\,\mu$m on a decade-long interstellar trip \citep{2016Icar..264..369P,2017ApJ...837....5H}. Experimental data for olivine particles with sizes of $0.3-1.2\,\mu$m accelerated towards an aluminum alloy (AlMg$_3$) at velocities of $3-7$\,km\,s$^{-1}$ show deepest craters of $1.64\,\mu$m \citep{MAPS:MAPS12338}. The crater depth is approximately linear with impact velocity between 2 and 8\,km\,s$^{-1}$ and can be extrapolated with hydrocode modeling to depths of $15\,\mu$m for $2\,\mu$m-sized particles at 10\,km\,s$^{-1}$ \citep{MAPS:MAPS1300}. Based on these estimates, it appears reasonable to design for a wall thickness of e.g., $150\,\mu$m (giving some safety margin) in aluminum. On average, such a shield will hold for $10^5$\,yrs based on the $10^{-9}\,$m\,yr$^{-1}$ erosion.

\subsection{Cosmic ray shielding}
In addition, the probe will be hit from all sides by cosmic rays. These are mostly hydrogen and helium nuclei accelerated to relativistic energies. At low Lorentz factors ($<0.3$) we can neglect the anisotropy caused by the probe's velocity. The flux can be approximated by monoenergetic nucleons with an average energy of 1 GeV and its rate is $\approx11$ nucleons cm$^{-2}$\,s$^{-1}$ \citep{2009AcAau..64..644S}.

Shielding against GeV particles is more effective when using light elements such as polyethylene (CH$_2$, $\rho=0.94\,\rm{g\,cm}^{-3}$) compared to aluminum because of fewer secondary cascades. Still, CH$_2$ shielding is relatively ineffective. A recent study by \citet{GUETERSLOH2006319} found that only 10\,\% of particles are stopped using a shield depth of 3\,cm. When increasing the shield to 10\,cm, the absorbed fraction increases to 30\,\%. The further relation between shield thickness and dose reduction is however not linear, and particle shields of close to 100\,\% effectiveness would require depths of several meters, which would result in prohibitively high masses. It appears preferable not to add extra shielding against cosmic rays, but instead invest the mass into error correction.

\subsection{Theoretical defects from cosmic ray hits}
\label{est}
For a cm-sized probe, most ($>90\,$\%) of cosmic rays will pass right through the payload, so that $P\approx0.1$ of the incoming particles produce primary recoils. Those that interact with the mass, however, will change the arrangement of the atoms in the crystal lattice. This radiation damage can be quantified with displacement theory. The primary knock-on atom moves through the lattice producing a collision cascade, which ends after the collision energy is smaller than the required minimum displacement energy.
To estimate the radiation damage, we take an average flux of monoenergetic GeV nucleons at a rate of $F_{\rm nucleons}\approx11$ cm$^{-2}$\,s$^{-1}$ \citep{2009AcAau..64..644S}. This flux is given per unit surface area. A solid physical body, e.g. a dice, has three spatial surfaces, so that a body of volume 1\,cm$^3$ collects $3 F_{\rm nucleons}$ per unit time and volume.

With these estimates, the number of cascades is 
$P \times 3 F_{\rm nucleons} 
\approx 0.1 \times 3 \times 11 \approx 3.3$\,s$^{-1}$\,cm$^{-3}$. Over the course of a year, this corresponds to $\approx10^8$ hits per cm$^3$. Each cascade displaces $<10^3$ atoms for primary hits of GeV energy in most materials \citep{Was2017}. The total effect is then $<10^{11}$ displaced atoms per year. This is a small fraction of the total number of $10^{23}$ atoms per cm$^3$.

Each displacement event will erase at most a few bits of information by modifying the chemical structure. Living organisms employ repair enzymes to cope with similar structural damage. A probe could in principle use nanobots but this would add mass and complexity. A better approach may be to pass the data-repair task onto the recipients; terrestrial molecular geneticists routinely use the BLAST \citep{Altschul1990} algorithm to reassemble corrupted information, though BLAST is a very general process and not restricted to reading DNA.

In total, unshielded cosmic rays can destroy only a small fraction of the payload even after long ($10^6$\,yrs) travel times. Most of the information can be protected using error correction schemes and/or duplication, which need to account for the fact that defects will occur in shapes of tunnels and clusters.

\subsection{Comparison to panspermia research}
Panspermia is the hypothesis that life is distributed by bodies such as asteroids. It makes for an interesting comparison, because DNA can store $5\times10^{20}$\,bits\,g$^{-1}$ \citep{2012Sci...337.1628C}. Stability is a function of temperature, and DNA decays quickly above 463\,K \citep{10.1089/dna.2013.2056}. At room temperature, its durability is estimated as 50\,yrs \citep{2013Natur.494...77G}. At cooler temperatures (286\,K), the half-life is estimated as 521 years inside bones \citep{10.1098/rspb.2012.1745}. In arctic ice ($\approx250$\,K), $>80\,$\% could be read after $8\times10^4$ yrs \citep{10.3402/polar.v34.25057}.

In a detailed review, the space radiation damage to different microbes was analyzed as a function of shield thickness and time. Some spores such as D. radiodurans were found to withstand a dose of $1{,}000\,$Gy, and $10^{-6}$ of the population would survive $10^6$\,yrs under minimum (cm) shielding \citep{2000Icar..145..391M}.

A probe carrying DNA would require a complete encasement of the DNA to avoid hydrolysis and vacuum damage. The survival time of bacteria due to DNA damage by hydrolysis is of the order $10^5$ years, based on available experimental evidence \citep{doi:10.1021/bi00567a010}. DNA decay by vacuum-caused damage occurs much faster than damage by hydrolysis or radiation, in tens of years \citep{1991OLEB...21..177D}. Unshielded micro-organisms are immediately killed by ultraviolet radiation in the solar system, and survive for $\approx10^5$ years with minimal shields. Outside of the solar system, the photon flux is $10^{-6}\times$ lower and can thus be neglected. Without repair, half-lifes of DNA onboard well-designed probes are $10^5-10^6$\,yrs.

\subsection{Probe geometry and structural integrity}
The ideal probe geometry is a long cylinder, because most of the shield mass is required for the front, $M_{\rm front} = \pi r^2 \rho \times (1 + 0.1d)\,$g at $v=0.1\,$c. Using graphite ($\rho=2.2\,\rm{g\,cm}^{-3}$) in a $r=1\,$cm wide probe to Alpha Cen at $v=0.1\,$c requires $M_{\rm front} \sim 9$\,g.

The mass of the side and bottom shields is $((\pi r^2) + (2 \pi r h)) \rho \,(2\times10^{-4}$\,g). If a given total shield mass is distributed evenly to the front and to the sides in a cylinder-shaped body, its length would be 150\,m (for $r=1$\,cm). Such a rod has a total shield mass of $\sim18\,$g and a volume of $\sim5\times10^4\,$cm$^{3}$. If the volume is filled with a storage mass of $\rho=2.2\,\rm{g\,cm}^{-3}$, the payload mass is $\sim26\,$kg. Consequently, the shield mass is only $\approx10^{-4}$ of the payload mass.

Such a very elongated rod might be unstable during a long-term (kyr, Myr) mission due to tumbling. A more conservative length to width ratio of $1:100$ (instead of $1:15{,}000$) would change the fraction of shield mass per payload only to $\approx10^{-2}$. Even in this conservative configuration, and with more generous shielding, the energy per bit does not change much. In the simple case where shield mass equals payload mass, the energy per bit increases by a factor of two compared to the values calculated in the following sections.

These calculations neglect the mass required for structural integrity at high accelerations, which might be much larger than the shield mass. If the structure is placed at the outside of the probe, it can serve double duty, so that the pure shield mass is even lower. 

Over time, dust hits on the hull will decrease stability and integrity of the structure. The shields will become perforated, and radiation damage will cause brittleness, so that cracks and fractures might form. A detailed analysis of the timescale of these effects is outside of the scope of this paper, but likely $>10^6\,$yrs.

\section{Information density}
\label{inscribed_matter}
A famous quote by \citet{Tanenbaum1989} highlights the data volume of inscribed matter: ``Never underestimate the bandwidth of a station wagon full of tapes hurtling down the highway.''.\footnote{Back in 1989, a common floppy-disk had a capacity of 1,474,560 bytes in the 3.5-inch format, which had a volume of $90\times94\times3.3$ mm, or 27.918\,cm$^3$. The world's largest road vehicles are probably to be found in Australia, with a ``K road truck'' pulling 6 wagons with a total capacity of $\sim750$m$^3$, fitting 27.16 million floppy disks, with a total capacity of $4\times10^{13}$ bytes, or 40 TB. This shows the exponential technological advances in storage technology; the same amount of data can today be saved on 80 Micro-SD cards of 512 GB each. These cards are tiny, so that within 27 years, we were able to shrink a gigantic road truck to the size of a golf ball.} Decades later, ``sneakernet'' is still used for high-latency high-bandwidth communication between computer systems. The same idea may be applied to interstellar communication.

Throughout this work, we neglect the energy required to encode the stored information, as it is small compared to transmission \citep{2004Natur.431...47R}.

\subsection{Theoretical limit of information per mass}
The physical maximum of information that can be stored within a spherical volume is given by the ``Bekenstein bound'' \citep{1972NCimL...4..737B,1973PhRvD...7.2333B,1974PhRvD...9.3292B}, which comes from the entropy of a black hole with the same surface area, and is $S_{\rm BH}=2\times10^{39}$\,bits\,g$^{-1}$. A solution without involving an inconvenient black hole has been found by scaling down solid-state flash memory to single electrons at $S_{\rm SS}=3\times10^{38}\,$bits\,g$^{-1}$ \citep{2013arXiv1309.7889K}.

\subsection{Nature and current technology}
DNA can store $5\times10^{20}$\,bits\,g$^{-1}$ \citep{2012Sci...337.1628C}. Typical proteins encode $2\times10^{21}$\,bits\,g$^{-1}$ \citep{2005q.bio....12025S}. Current technology offers $10^{18}\,$bits\,g$^{-1}$ in magnetic hard disks on a thinned 10\,nm cobalt layer, and similar values for solid-state disks. Practical disks are much heavier because of durability, read-/write devices, and enclosures.

\subsection{Practical limit of information per mass}
We now estimate a (conservative, robust) information density limit with ``Earth 2018'' technology.

Let a tape consist of a long chain of carbon atoms terminated by e.g., a \ce{-CCl_3} group at one end and a \ce{-C(Cl_3)_3} group at the other so that the ends of the tape can be distinguished. The actual terminators used could be almost anything, as long as they are distinguishable. Each intervening carbon atom can be bound to a pair of univalent radicals. For simplicity assume that only two different radicals are used, denoted by \ce{-A} and \ce{-B}. A short tape might then be:

\vspace{0.25cm}
\chemfig{CCl_3-C(-[2]A)(-[6]A)-C(-[2]B)(-[6]A)-C(-[2]B)(-[6]B)-C[CCl_3]_3}
\vspace{0.25cm}

An order of magnitude gain in storage density could be reached by encoding more bits per a.m.u. by using e.g., \ce{-H}, \ce{-D}, \ce{-CH_3}, \ce{-CH_2D}, \ce{-CHD_2} and \ce{-CD_3} radicals. Such an encoding scheme would allow for 23 states to be encoded in an average of 35\,a.m.u each, which corresponds to $35 \times \rm{log}_2(23) \approx 7.7\,$a.m.u per bit. For simplicity and increased stability, we continue with perfluorinated alkane as an example.

Each carbon atom in the chain can have one of three substitution patterns (the formulae \ce{-CAB-} and \ce{-CBA-} are structurally and chemically identical) and can store one trinary digit (tit) or $\rm{log}_2(3) \approx 1.6$ bits.

The lowest mass choices for the A and B radicals are -H and -D with masses of 1 and 2\,a.m.u. respectively. The atomic mass of carbon is 12\,a.m.u. so assuming equal populations of each combination the average mass for this case is 15\,a.m.u per tit. Reading and writing the tape would be easier, from a chemical point of view, if A were hydrogen and B were methyl \ce{(-CH_3)} which has a mass of $15\,$a.m.u. The average mass now becomes 28\,a.m.u per tit. Long term chemical stability suggests that rather than hydrogen we should use fluorine atoms, which have a mass of 19\,a.m.u, so the average mass per tit rises to 100\,a.m.u per tit. 

One a.m.u is equal to $1.66\times 10^{-27}$\,kg, so that the described encoding schemes allow for $S_{\rm chem}=6\times10^{23}$\,bits\,g$^{-1}$, or a magnitude less or more depending on the desired stability.

Compared to DNA, perfluorocarbon chemistry provides higher stability at low temperature and in the absence of vicious reagents by orders of magnitude. Its chemistry stops $\lesssim500\,$K, an advantage for the short periods of high temperatures during a laser-pushed launch \citep{2016arXiv160401356L,2017AJ....153..277K} and the close stellar encounter of a photogravimagnetic deceleration \citep{2017ApJ...835L..32H,2017AJ....154..115H,2017arXiv171105856F}.

The enclosure of the probe, structural components, and overhead for error detection and correction codes will make the information payload a fraction of the total probe mass, e.g. 20\%. We use $S_{\rm base}=10^{23}$\,bits\,g$^{-1}$ in the following sections, and the reader is welcome to favor other values. As Richard \citet{128057} noted: ``There's plenty of room at the bottom''.

\section{Transportation}
It is still debated whether there is a lower limit on the minimal energy required per transmitted bit \citep{1988ApPhL..52.2191P,1996Sci...272.1914L}. In practice, however, interstellar communication needs to leave our solar system, and thus massive particles require a minimum speed of $\approx20\,$km\,s$^{-1}$, the solar system escape velocity. Mass-less particles have comparable energy limits because of thermal noise, as pointed out by \citet{holevo1973bounds,2014NaPho...8..796G}.

A key advantage of inscribed matter communications is that its energy requirements are independent of distance, instead they are a function of velocity. In contrast, a particle beam's width widens with distance, making the receiver flux level an inverse quadratic function of distance. While mass-less particles are cheap (there are $5\times10^{15}\,$ photons per Joule at $\lambda=$\,nm), it gives an advantage to matter over photons for larger distances. In any comparison between matter and particle communication there is a distance $d_{\rm 0}$ after which probes are more energy efficient for a given technology (but require longer wait times).

Theoretically, the kinetic energy invested into accelerating a mass can (almost) be recovered during its deceleration, making such communications extremely energy efficient \citep{1998RSPSA.454..305L}. In practice, this would require the construction of megastructures in space and is neglected here: We will instead assume that the kinetic energy is required twice.

\subsection{Potential transportation scheme}
A recent proposal by the ``Breakthrough Initiatives'' \citep{2016arXiv160401356L,2017Natur.542...20P} suggests that near future technology can launch 1\,g payloads with a cruise velocity of $0.2\,$c using strong lasers. Such a probe would arrive at the nearest star, $\alpha$\,Cen, after about 20 years, corresponding to a data rate of e.g. $10^{18}\,$bits over 20 years, or $10^{9}\,$bits per second (Gbit/s). Launching probes regularly, e.g. one per day, increases the effective data rate by orders of magnitude.

\subsection{Energy requirements per bit of inscribed matter}
Accelerating a probe to velocities $v$ close to the speed of light $c$ requires a relativistic treatment. For slow probes $v \ll c$ we can approximate the Lorentz factor as $L=1$ with an error of $\approx5\,$\% ($\approx15\,$\%, $\approx100\,$\%) at 0.3\,c (0.5\,c, 0.9\,c), or calculate

\begin{equation}
L = \frac{1}{\sqrt{1 - v^2/c^2}} .
\end{equation}

The kinetic energy of an object is

\begin{equation}
E_{\rm kin} = \frac{1}{2}mv^2
\end{equation}

where $m$ is the object's mass (in kg), and $v$ its velocity (in m\,s$^{-1}$). With $\eta$ as the propulsion efficiency factor for launch and deceleration, the capacity of the channel is

\begin{equation}
C_{\rm rel} = \eta\, S\, L^{-1}\, v^{-2} \,\,(\rm {bits\,J^{-1}}).
\end{equation}

With $S_{\rm base}=10^{23}$\,bits\,g$^{-1}$ and $v=0.1\,c$ we get an energy efficiency of $\approx10^{11}$\,bits per Joule, where we can neglect the small Lorentz factor ($L=1.005$).

\section{Discussion}

\subsection{Comparison to photon communications}
We can compare matter to the optimal photon communication at $\lambda\approx$\,nm where the number of received particles is \citep{2017arXiv171105761H}

\begin{equation}
\gamma \sim 
\left(\frac{d}{1\,{\rm pc}} \right)^{-2} 
\left(\frac{D_{\rm t}}{1\,{\rm m}} \right)^{2}
\left(\frac{D_{\rm r}}{1\,{\rm m}} \right)^{2}
\left(\frac{P}{1\,{\rm W}} \right)
\,\,{\rm (s^{-1})}
\end{equation}

with $d$ as the distance, $D_{\rm t}$ and $D_{\rm r}$ as transmitter and receiver apertures, and $P$ as the transmitter power. Using a conservative capacity of one bit per photon \citep{2017arXiv171205682H}, setting $D=D_{\rm t}=D_{\rm r}$, and dividing by power, we can write the capacity of the channel as a function of wavelength and in units of bits per energy:

\begin{equation}
C_{\gamma} = 
\eta\,
\left(\frac{d}{1\,{\rm pc}} \right)^{-2} 
\left(\frac{\lambda}{1\,{\rm nm}} \right)^{-1} 
\left(\frac{D}{1\,{\rm m}} \right)^{4}
\,\,(\rm {bits\,J^{-1}}).
\end{equation}

We can set $C_{\gamma}=C_{\rm rel}$ to match the energy efficiency of $10^{11}$\,bits per Joule as in the inscribed matter case. For $d=1.3\,$pc, we require $D = 1\,$km at X-ray energy, which is large. For optical lasers ($\lambda=\mu$m) and microwaves ($\lambda=0.1\,$m), the apertures would need to be larger by $10\times$ and $100\times$, respectively.

For a constant wavelength, the apertures must increase with distance $d$ as $D \propto d^{1/2}$. For a constant $v=0.1\,$c, the apertures increase with distances $d=10\,$pc ($100$\,pc, $1{,}000\,$pc) to $D=3\,$km  (9\,km, 27\,km).

For low probe speeds ($v \ll 0.01\,$c), aperture sizes for energy equivalence become implausibly high. In this regime, inscribed matter is more energy efficient by many orders of magnitude.

\subsection{Limits on energy and aperture size}
In the real world, the amount of energy used to accelerate (and decelerate) the probe is limited, as is the beamer size. As an example, ``Breakthrough Starshot'' proposes to use a 100\,GW laser with a 10\,km aperture, sufficient to accelerate 1\,g to $v=0.2\,$c. At first approximation, mass and travel time can be traded. At a slower velocity of e.g., $v=0.002\,$c, the mass could be 100\,g for a travel time to $\alpha\,$Cen of $2{,}000\,$yrs, which is acceptable with respect to decay. Practical limits on power and aperture impose a limit on mass and velocity. Without megastructures and extreme power levels, masses will plausibly be limited to the kg range.

\subsection{Comparison to photon communications through stellar gravitational lenses}
Alternatively, meter-scale telescopes could be placed in the gravitational lenses of both stars, achieving an equivalent of 75\,km apertures on the receiving legs only \citep{2017arXiv170605570H}. Using meter-size transmitters, photon communication is more efficient over 1.3\,pc compared to a probe with $v=0.1\,$c, and achieves parity at kpc distances.

\subsection{Decoupling of times}
As explained in the introduction, inscribed matter probes can be employed for different purposes. One scenario is to send a library onboard a sentinel with the wishful thinking that something may eventually evolve the capability to read it.

An active ``sneakernet'', in contrast, presupposes that a channel set-up protocol has already been established before the first probe is sent. Then, a large bulk of data can be transmitted relatively cheaply without clogging up a bosonic network connection. This scenario resembles today's occasional transfers of large amounts of data via traditional mail channels, e.g. on disks.

Sending a sentinel also removes the requirement of synchronicity. It allows the sender to decouple the time required for information preparation and transmission. For example, a probe (or many probes) can be assembled in parallel, perhaps consisting of many separate storage chips. Each probe can be send with a catapult (or photon beamer, etc.) in a relatively short time (minutes). When we compare a g-mass probe (with $10^{23}\,$bits of information), a launch to 0.2\,c would take $\approx2$ minutes, compared to a multi-year transmission campaign for the same amount of data. The quantity of data per unit time can thus be much larger than with a bosonic transmitter, and can be targeted at many different receivers, as long as energy is not the bottleneck.

In a bosonic channel, sender and receiver must spend the same amount of time, and the same time (plus light travel time) for communication. If the particles are not received, the information is lost forever. As an analogy, the bosonic channel would be a live TV broadcast, while the probe would represent a letter waiting in your mailbox.

\subsection{Probe strategies}
In principle, and in contrast to the communication strategies discussed in the previous section, no network needs to be established before sending a probe. As an example, the sender of the probe may estimate a relevant chance that a technological species exists, or may evolve in the future, in a target system. The sender can launch a probe with a trajectory to be captured at the destination system e.g., via three-body gravitational interactions involving the probe, a gas giant such as Jupiter, and the star. Interstellar comets, similar to Oumuamua \citep{
2017Natur.552..378M}, are regularly captured by the solar system ``fishing net'' \citep{2018arXiv180110254L}. Alternatively, deceleration with a light sail is possible from a few percent the speed of light, using a photogravimagnetic manoeuvre \citep{2017ApJ...835L..32H,2017AJ....154..115H,2017arXiv171105856F}. The probe can then be landed on a moon, parked in a stable orbit around Lagrangian points, or may be actively controlled e.g., by an on-board AI.

The efficiency of probes compared to photons is strongly related to the communication distance. If the average distance between species is short, photons might be preferred in many cases. In regions with high stellar densities, such as globular clusters or the galactic center, the average distance between stars is an order of magnitude shorter \citep{2013ApJ...774..151M}. If these places are suitable for technological life, they may be dominated by photon communications. In the outskirts of the galaxy, the opposite may be true.

\section{Conclusion}
Probes which carry inscribed matter have higher latency compared to photon communications; realistic cases are a factor of $10\times$ for $v=0.1\,$c. If this delay is acceptable for both sides, inscribed matter is favorable for cases where large amounts of data ($>10^{23}$\,bits) shall be transmitted over a comparably short time (centuries). Its energy efficiency (per bit) increases with distance compared to photons, and is equivalent to photon communications with large (km) apertures over pc distances. Practical limits due to durability arise for travel times $>10^6\,$yrs, which covers all of the galaxy at $v=0.1\,$c, but makes intergalactic communication prohibitive. Stellar gravitational lensing at the receiver side is more efficient out to kpc, limiting inscribed matter to scenarios such as the long-term deposition of artifacts inside a stellar system.

\pagebreak
\section{References}
\bibliographystyle{elsarticle-harv}
\bibliography{references_elsevier}
\end{document}